\documentclass[twocolumn,english,aps,twocolumn]{revtex4}
\usepackage[T1]{fontenc}
\usepackage[latin9]{inputenc}
\usepackage{color}
\usepackage{textcomp}
\usepackage{amsthm}
\usepackage{amsmath}
\usepackage{graphicx}
\usepackage{amssymb}
\usepackage{esint}

\makeatletter

\newcommand{\lyxmathsym}[1]{\ifmmode\begingroup\def\b@ld{bold}
  \text{\ifx\math@version\b@ld\bfseries\fi#1}\endgroup\else#1\fi}

\newcommand{\lyxdot}{.}

\@ifundefined{textcolor}{}
{%
 \definecolor{BLACK}{gray}{0}
 \definecolor{WHITE}{gray}{1}
 \definecolor{RED}{rgb}{1,0,0}
 \definecolor{GREEN}{rgb}{0,1,0}
 \definecolor{BLUE}{rgb}{0,0,1}
 \definecolor{CYAN}{cmyk}{1,0,0,0}
 \definecolor{MAGENTA}{cmyk}{0,1,0,0}
 \definecolor{YELLOW}{cmyk}{0,0,1,0}
 }

\makeatother

\usepackage{babel}

\begin{document}

\title{Characterization of a two-transmon processor with individual single-shot
qubit readout }

\author{A. Dewes$^{1}$, F. R. Ong$^{1}$, V. Schmitt$^{1}$, R. Lauro$^{1}$,
N. Boulant$^{2}$, P. Bertet$^{1}$, D. Vion$^{1}$, and D. Esteve$^{1}$}

\affiliation{$^{1}$Quantronics group, Service de Physique de l'État Condensé
(CNRS URA 2464), IRAMIS, DSM, CEA-Saclay, 91191 Gif-sur-Yvette, France }

\affiliation{$^{2}$LRMN, Neurospin, I2BM, DSV , 91191CEA-Saclay, 91191 Gif-sur-Yvette,
France }
\begin{abstract}
We report the characterization of a two-qubit processor implemented
with two capacitively coupled tunable superconducting qubits of the
transmon type, each qubit having its own non-destructive single-shot
readout. The fixed capacitive coupling yields the $\sqrt{iSWAP}$
two-qubit gate for a suitable interaction time. We reconstruct by
state tomography the coherent dynamics of the two-bit register as
a function of the interaction time, observe a violation of the Bell
inequality by 22 standard deviations after correcting readout errors,
and measure by quantum process tomography a gate fidelity of 90\%.
\end{abstract}
\maketitle
\textcolor{black}{Quantum information processing is one of the most
appealing ideas for exploiting the resources of quantum physics and
performing tasks beyond the reach of classical machines \cite{Nielsen Chuang}.
Ideally, a quantum processor consists of an ensemble of highly coherent
two-level systems, the qubits, that can be efficiently reset, that
can follow any unitary evolution needed by an algorithm using a universal
set of single and two qubit gates, and that can be readout projectively.
In the domain of electrical quantum circuits \cite{QubitsSupra},
important progress \cite{DiCarlo2qb,Yamamoto,Bialczak,IBM Steffen,DiCarlo3qb}
has been achieved recently with the operation of elementary quantum
processors based on different superconducting qubits. Those based
on transmon qubits \cite{DiCarlo2qb,DiCarlo3qb,Transmon Koch,TransmonSchreier}
are well protected against decoherence but embed all the qubits in
a single resonator used both for coupling them and for joint readout.
Consequently, individual readout of the qubits is not possible and
the results of a calculation, as the Grover search algorithm demonstrated
on two qubits \cite{DiCarlo2qb}, cannot be obtained by running the
algorithm only once. Furthermore, the overhead for getting a result
from such a processor without single-shot readout but with a larger
number of qubits overcomes the speed-up gain expected for any useful
algorithm. The situation is different for processors based on phase
qubits \cite{Yamamoto,Bialczak,RezQ}, where the qubits are more sensitive
to decoherence but can be read individually with high fidelity, although
destructively. This significant departure from the wished scheme can
be circumvented, when needed, since a destructive readout can be transformed
into a non-destructive one at the cost of adding one ancilla qubit
and one extra two-qubit gate for each qubit to be read projectively.
Moreover, energy release during a destructive readout can result in
a sizeable cross-talk between the readout outcomes, which can also
be solved at the expense of a more complex architecture \cite{Ansmann Bell,RezQ}. }

\begin{figure}[t]
\includegraphics[width=8cm]{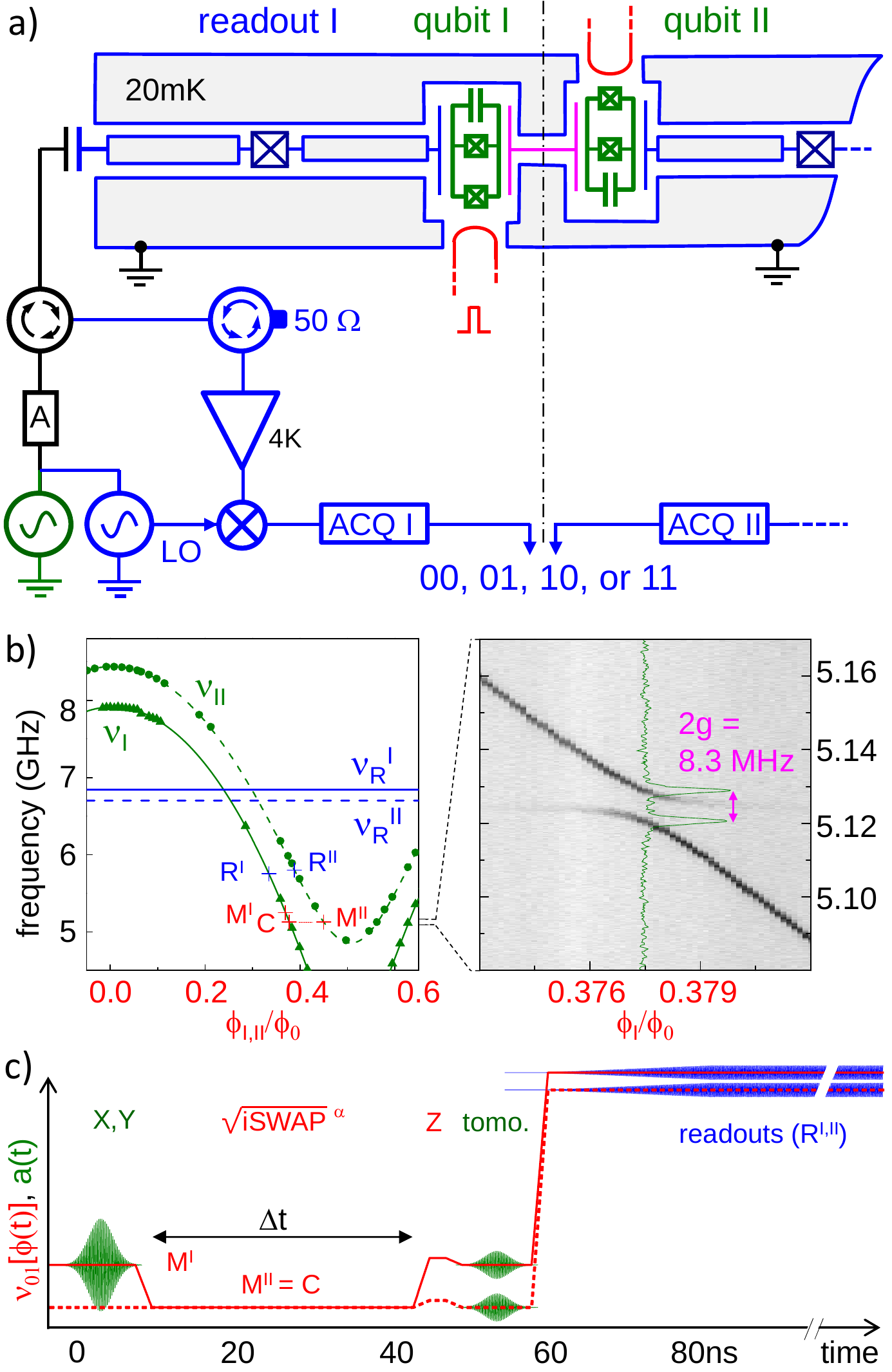}\caption{(a): circuit schematics \textcolor{black}{of the experiment with qubits
in green and readout circuits in grayed blue}. (b) Left panel: Spectroscopy
of the sample showing the resonator frequencies $\nu_{\mathrm{R}}^{\mathrm{\mathrm{I},II}}$
(horizontal lines), and the measured (disks, triangles)\textcolor{black}{{}
and fitted (lines) }qubit frequencies $\nu_{\mathrm{I,II}}$ as a
function of their flux bias $\phi_{\mathrm{I,II}}$ when the other
qubit is far detuned. Right panel: Spectroscopic anticrossing of the
two qubits revealed by the 2D plot of $p_{\mathrm{01}}+p_{\mathrm{10}}$
as a function of the probe frequency and of $\phi_{\mathrm{I}}$,
at $\nu_{\mathrm{II}}=5.124\,\mathrm{GHz}$. (c) Typical pulse sequence
including $X$ or $Y$ rotations, a $\sqrt{iSWAP}{}^{\alpha}$gate,
$Z$ rotations, and tomographic and readout pulses. Microwave pulses
$a(t)$ for qubit (green) and for readout (blue) are drawn on top
of the $\nu_{\mathrm{I,II}}(\phi)$ dc pulses (red lines).}

\label{fig1}%
\end{figure}
\textcolor{black}{In this work, we operate a new architecture that
comes closer to the ideal quantum processor design than the above-mentioned
ones. Our circuit is based on frequency tunable transmons that are
capacitively coupled. Although the coupling is fixed, the interaction
is effective only when the qubits are on resonance, which yields the
$\sqrt{iSWAP}$ universal gate for an adequate coupling duration.
Each qubit is equipped with its own non-destructive single-shot readout
\cite{JBA Siddiqi,JBA Mallet} and the two qubits can be read with
low cross-talk. In order to characterize the circuit operation, we
reconstruct the time evolution of the two-qubit register density matrix
during the resonant and coherent exchange of a single quantum of excitation
between the qubits by quantum state tomography. Then, we prepare a
Bell state with concurrence 0.85, measure the CHSH entanglement witness,
and find a violation of the corresponding Bell inequality by 22 standard
deviations. We then characterize the $\sqrt{iSWAP}$ universal gate
operation by determining its}\textcolor{magenta}{{} }\textcolor{black}{process
map with quantum process}\textcolor{red}{{} }\textcolor{black}{tomography
\cite{Nielsen Chuang}. We find a gate fidelity of 90\% due to qubit
decoherence and systematic unitary errors. }

The circuit implemented is schematized in Fig.\ref{fig1}a: the \textcolor{black}{coupled
}qubits with their respective \textcolor{black}{control and readout
sub-circuits are} fabricated on a Si chip (see supplementary information
S1). The chip is cooled down to $20\,\mathrm{mK}$ in a dilution refrigerator
and connected to room temperature sources and measurement devices
by attenuated and filtered control lines and by two measurement lines
equipped with cryogenic amplifiers. Each transmon $j=I,\, II$ is
a capacitively shunted SQUID characterized by its Coulomb energy $E_{\mathrm{C}}^{\mathrm{j}}$
for a Cooper pair, the asymmetry $d_{\mathrm{j}}$ between its two
Josephson junctions, and its total effective Josephson energy $E_{\mathrm{J}}^{\mathrm{j}}(\phi_{j})=E_{\mathrm{J}}^{\mathrm{j}}\left|\mathrm{cos}(x_{\mathrm{j}})\right|\sqrt{1+d_{\mathrm{j}}^{2}\mathrm{tan^{2}}(x_{\mathrm{j}})}$,
with $x_{\mathrm{j}}=\pi\phi_{\mathrm{j}}/\phi_{0}$, $\phi_{0}$
the flux quantum, and $\phi_{\mathrm{j}}$ the magnetic flux through
the SQUIDs induced by two local current lines with a $0.5\,\mathrm{GHz}$
bandwidth. The transition frequencies $\nu_{\mathrm{j}}\simeq\sqrt{2E_{\mathrm{C}}^{\mathrm{j}}E_{\mathrm{J}}^{\mathrm{j}}}/h$
between the two lowest energy states $\left|0\right\rangle _{j}$
and $\left|1\right\rangle _{j}$ can thus be tuned by $\phi_{\mathrm{j}}$.
The qubits are coupled by a capacitor with nominal value $C_{c}\simeq0.13\,\mathrm{fF}$
and form a register with Hamiltonian $H=h\left(-\nu_{\mathrm{I}}\sigma_{\mathrm{z}}^{\mathrm{I}}-\nu_{\mathrm{II}}\sigma_{\mathrm{z}}^{\mathrm{II}}+{\color{black}{\color{red}{\color{black}2}}}g\sigma_{\mathrm{y}}^{\mathrm{I}}\sigma_{\mathrm{y}}^{\mathrm{II}}\right)/2$.
Here $h$ is the Planck constant, $\sigma_{\mathrm{x,y,z}}$ are the
Pauli operators, ${\color{red}{\color{black}2}}g=\sqrt{E_{\mathrm{C}}^{\mathrm{I}}E_{\mathrm{C}}^{\mathrm{II}}\nu_{\mathrm{I}}\nu_{\mathrm{II}}}/E_{\mathrm{Cc}}\ll\nu_{\mathrm{I,II}}$
is the coupling frequency, and $E_{\mathrm{Cc}}$ the Coulomb energy
of a Cooper pair on the coupling capacitor. The two-qubit gate is
defined in the uncoupled basis $\{\left|uv\right\rangle \}\equiv\{\left|0\right\rangle _{\mathrm{I}},\left|1\right\rangle _{\mathrm{I}}\}\otimes\{\left|0\right\rangle _{\mathrm{II}},\left|1\right\rangle _{\mathrm{II}}\}$,
at a working point $M_{\mathrm{I,II}}$ where the qubits are sufficiently
detuned ($\nu_{\mathrm{II}}-\nu_{\mathrm{I}}\gg{\color{red}{\color{black}2}}g$)
to be negligibly coupled. Bringing them on resonance at a frequency
$\nu$ in a time much shorter than $1/2g$ but much longer than 1/$\nu$,
and keeping them on resonance during a time $\Delta t$, one implements
an operation $\Theta_{\mathrm{I}}.\Theta_{\mathrm{II}}.\sqrt{iSWAP}^{\,(8g\Delta t)}$,
which is the product of the\[
\sqrt{iSWAP}=\left(\begin{array}{cccc}
1 & 0 & 0 & 0\\
0 & 1/\sqrt{2} & -i/\sqrt{2} & 0\\
0 & -i/\sqrt{2} & 1/\sqrt{2} & 0\\
0 & 0 & 0 & 1\end{array}\right)\]
gate to an adjustable power and of two single qubit phase gates $\Theta_{\mathrm{j}}=\mathrm{exp\left(i\theta_{j}\sigma_{z}^{j}{\color{red}{\color{black}/2}}\right)}$
accounting for the dynamical phases $\theta_{\mathrm{j}}=\int{\color{red}{\color{black}2}}\pi(\nu-\nu_{\mathrm{j}})dt$
accumulated during the coupling. The exact $\sqrt{iSWAP}$ gate can
thus be obtained by choosing $\Delta t={\color{red}{\color{black}1}}/8g$
and by applying a compensation rotation $\Theta_{\mathrm{j}}^{-1}$
to each qubit afterward.

For readout, each qubit is capacitively coupled to its own \textcolor{black}{$\lambda/2$
}coplanar waveguide resonator \textcolor{black}{with frequency $\nu_{\mathrm{R}}^{\mathrm{j}}$
and quality factor $Q_{\mathrm{j}}$. This resonator is made non linear
with a Josephson junction and }is operated as a Josephson bifurcation
amplifier, as explained in detail in \cite{JBA Mallet}. The homodyne
measurement \textcolor{black}{(see Fig.\ref{fig1}a) }of two microwave
pulses simultaneously applied to and reflected from the resonators
yields a two-bit outcome $uv$ that maps with a high fidelity the
state $\left|uv\right\rangle $ on which the register is projected;
the probabilities $p_{\mathrm{uv}}$ of the four possible outcomes
are determined by repeating the same experimental sequence a few $10^{4}$
times. Single qubit rotations $u\left(\theta\right)$ by an angle
$\theta$ around an axis $\overrightarrow{u}$ of the XY plane of
the Bloch sphere are obtained by applying Gaussian microwave pulses
through the readout resonators, with frequencies $\nu_{\mathrm{j}}$,
phases $\varphi_{\mathrm{j}}=(\overrightarrow{X},\overrightarrow{u})$,
and calibrated area $A_{\mathrm{j}}\varpropto\theta$. Rotations around
Z are obtained by changing temporarily $\nu_{\mathrm{I,II}}$ with
dc pulses on the current lines.

The sample is first characterized by spectroscopy (see Fig.\ref{fig1}b)
and a fit of the transmon model to the data yields the sample parameters
(see S2). The working points where the qubits are manipulated $(M^{\mathrm{I,II}}$),
resonantly coupled (C), and read out $(R^{\mathrm{I,II}}$) are chosen
to yield sufficiently long relaxation times $\sim0.5\,\text{\textmu s}$
during gates, negligible residual coupling during single qubit rotations
and readout, and best possible fidelities at readout. Figure \ref{fig1}b
shows these points as well as the spectroscopic anticrossing of the
two qubits at point $C$, where ${\color{red}{\color{black}2}}g=8.3\;\mathrm{MHz}$
in agreement with the design value of $C_{\mathrm{c}}$. Then, readout
errors are characterized at $R^{\mathrm{I,II}}$ (see Fig. S3.1):
In a first approximation, the errors are independent for the two readouts
and are of about 10\% and 20\% when reading $\left|0\right\rangle $
and $\left|1\right\rangle $ respectively. This limited fidelity results
for a large part from energy relaxation of the qubits at readout.
In addition we observe a small readout cross talk, i.e. a variation
of up to 2\% in the probability of an outcome of readout $j$ depending
on the state of the other qubit. All these effects are calibrated
by measuring the four $p_{\mathrm{uv}}$ probabilities for each of
the four $\left|uv\right\rangle $ states, which allows us to calculate
a $4\times4$ readout matrix $\mathcal{R}$ linking the $p_{\mathrm{uv}}$'s
to the $\left|uv\right\rangle $ populations.

\begin{figure}[h]
\includegraphics[width=8cm]{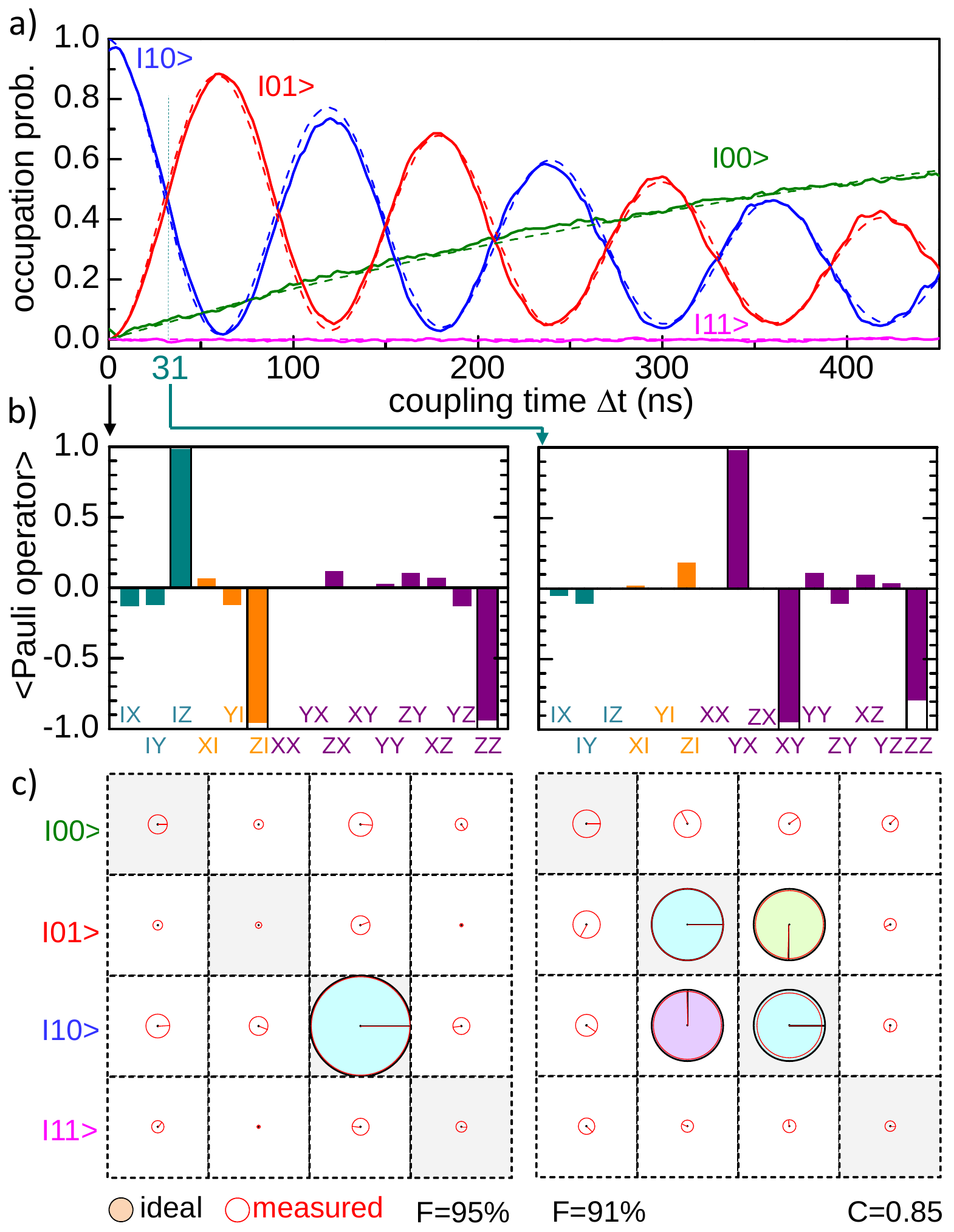}\caption{Coherent swapping of a single excitation between the qubits. (a) Experimental
(solid lines) and fitted (dashed lines) occupation probabilities of
the four computational states $\left|00\right\rangle ..\left|11\right\rangle $
as a function of the coupling duration. No $Z$ or tomographic pulses
are applied here. (b,c) State tomography of the initial state (left)
and of the state produced by the $\sqrt{iSWAP}$ gate (right). (b)
Ideal (empty bars) and experimental (color filling) expected values
of the 15 Pauli operators $XI,..,ZZ$. (c) Corresponding ideal (color
filled black circles with pointer) and experimental (red circle and
pointer) density matrices, as well as fidelity $F$ and concurence
$C$. Each matrix element is represented by a circle with an area
proportional to its modulus (diameter = cell size for unit modulus)
and a phase pointer giving its argument.}

\label{fig2}%
\end{figure}

Repeating the pulse sequence shown in Fig.\ref{fig1}c at $M^{\mathrm{I}}=5.247\,\mathrm{GHz}$,
$M^{\mathrm{II}}=C=5.125\,\mathrm{GHz}$, $R^{\mathrm{I}}=5.80\,\mathrm{GHz}$,
$R^{\mathrm{II}}=5.75\,\mathrm{GHz}$, and applying the readout corrections
$\mathcal{R}$, we observe the coherent exchange of a single excitation
initially stored in qubit $I$. We show in Fig.\ref{fig2} the time
evolution of the measured $\left|uv\right\rangle $ populations, in
fair agreement with a prediction obtained by integration of a simple
time independent Liouville master equation of the system, involving
the independently measured relaxation times $T_{1}^{\mathrm{I}}=436\,\mathrm{ns}$
and $T_{1}^{\mathrm{II}}=520\,\mathrm{ns}$, and two independent effective
pure dephasing times $T_{\mathrm{\varphi}}^{\mathrm{I}}=T_{\mathrm{\varphi}}^{\mathrm{II}}=2.0\,\mu s$
as fitting parameters. Tomographic reconstruction of the register
density matrix $\rho$ is obtained by measuring the expectation values
of the 15 two-qubit Pauli operators $\left\{ P_{\mathrm{k}}\right\} =\left\{ XI,..,ZZ\right\} $,
the $X_{\mathrm{j}}$ and $Y_{\mathrm{j}}$ measurements being obtained
using tomographic pulses $\overrightarrow{Y_{\mathrm{j}}}\left(-90\text{\textdegree}\right)$
or $\overrightarrow{X_{\mathrm{j}}}\left(90\text{\textdegree}\right)$
just before readout. The $\rho$ matrix is calculated from the Pauli
set by global minimization of the Hilbert-Schmidt distance between
the possibly non-physical $\rho$ and all physical (i.e. positive-semidefinite)
$\rho's$. This can be done at regular interval of the coupling time
to produce a movie of $\rho\left(\Delta t\right)$ (see supplementary
on line material) showing the swapping of the $\left|10\right\rangle $
and $\left|01\right\rangle $ populations at frequency ${\color{red}{\color{black}2}}g$,
the corresponding oscillation of the coherences, as well as the relaxation
towards $\left|00\right\rangle $. Figure \ref{fig2} shows $\left\{ \left\langle P_{\mathrm{k}}\right\rangle \right\} $
and $\rho$ only at $\Delta t$=$0\,\mathrm{ns}$ and after a $\sqrt{iSWAP}$
obtained at $\Delta t$=$31\,\mathrm{ns}$ with $\Theta_{\mathrm{j}}^{-1}$rotations
of $\theta_{\mathrm{I}}\simeq-65\text{\textdegree}$ and $\theta_{\mathrm{II}}\simeq+60\text{\textdegree}$.
The fidelity $F=\left\langle \psi_{\mathrm{id}}\right|\rho\left|\psi_{\mathrm{id}}\right\rangle $
of $\rho$ with the ideal density matrices $\left|\psi_{\mathrm{id}}\right\rangle \left\langle \psi_{\mathrm{id}}\right|$
are 95\% and 91\%, respectively, and are limited by errors on the
preparation pulse, statistical noise, and relaxation.

\begin{figure}
\includegraphics[width=8cm]{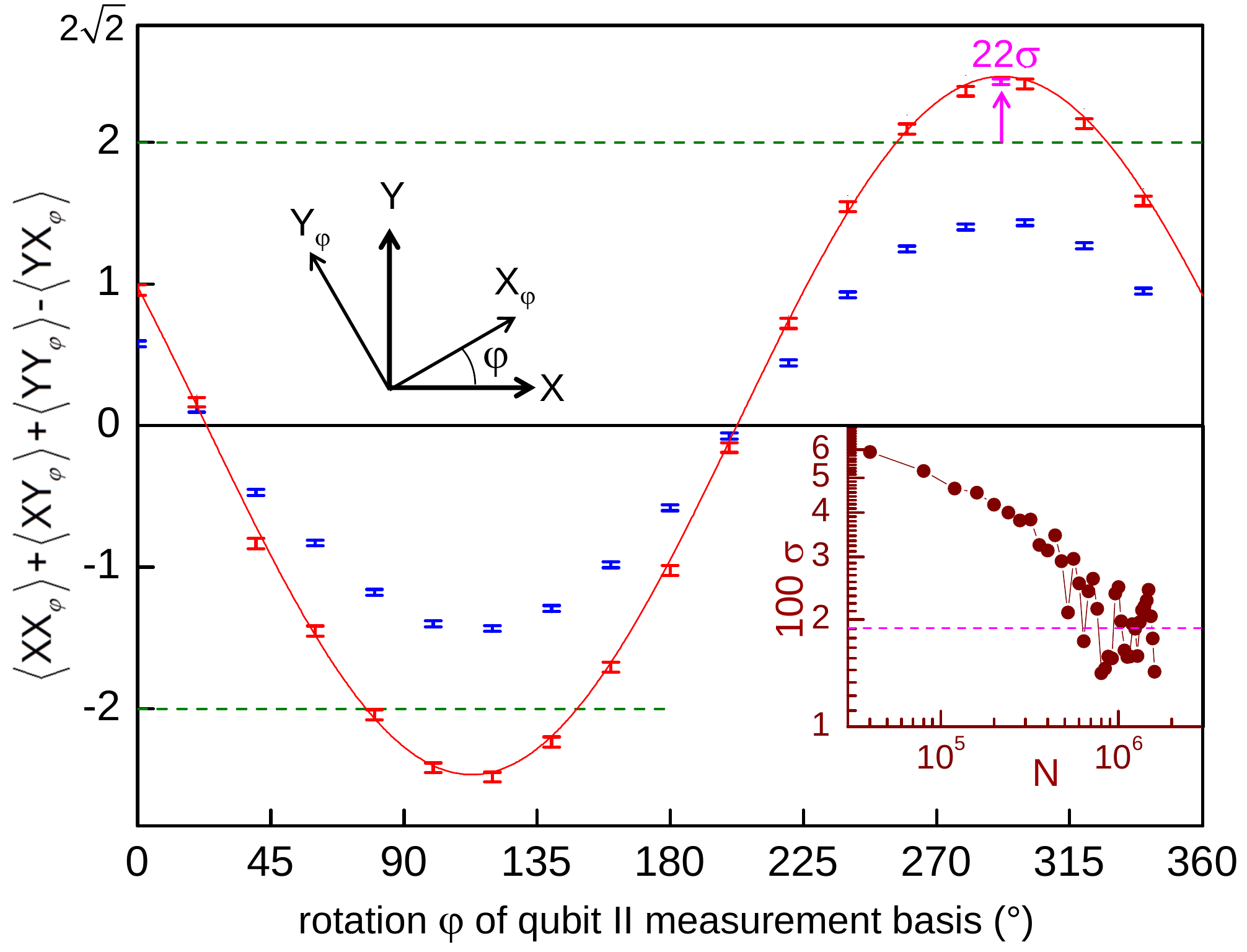}

\caption{Test of the CHSH-Bell inequality on a $\left|10\right\rangle +e^{\mathrm{i\psi}}\left|01\right\rangle $
state by measuring the qubits along $X^{\mathrm{I}}$or $Y^{\mathrm{I}}$
and $\mathrm{X_{\mathrm{\varphi}}^{II}}$ or $Y_{\mathrm{\varphi}}^{\mathrm{II}}$
(see top-left inset)\textcolor{black}{,}\textcolor{red}{{} }\textcolor{black}{respectively}.
Blue (resp. red) error bars are the experimental CHSH entanglement
witness determined from the raw (resp. readout errors corrected) measurements
as a function of the angle $\varphi$ between the measuring basis,
whereas solid line is a fit using $\psi$ as the only fitting parameter.
Height of error bars is $\pm$ one standard deviation $\sigma(N)$
(see bottom-right inset), with $N$ the number of sequences per point.
Note that averaging beyond $N=10^{6}$ does not improve the violation
because of a slow drift of $\varphi$.}

\label{fig3}%
\end{figure}

To quantify in a different way our ability to entangle the two qubits,
we prepare a Bell state $\left|10\right\rangle +e^{\mathrm{i\psi}}\left|01\right\rangle $
\textcolor{black}{(with }$\psi=\theta_{\mathrm{II}}-\theta_{\mathrm{I}}$)\textcolor{black}{{}
using the pulse sequence of Fig.\ref{fig1}c with $\Delta t=31\,\mathrm{ns}$
and no $\Theta_{j}^{-1}$rotations}, and measure the CHSH entanglement
witness $\left\langle XX_{\mathrm{\varphi}}\right\rangle +\left\langle XY_{\mathrm{\varphi}}\right\rangle +\left\langle YY_{\mathrm{\varphi}}\right\rangle -\left\langle YX_{\mathrm{\varphi}}\right\rangle $
as a function of the angle $\varphi$ between the orthogonal measurement
bases of qubit $I$ and $II$. Figure \ref{fig3} compares the results
obtained with and without correcting the readout errors, with what
is theoretically expected from the decoherence parameters indicated
previously: unlike in \cite{Ansmann Bell} and because of a readout
contrast limited to $70-75\,\%$, the witness does not exceed the
classical bound of 2 without correcting the readout errors. After
correction, it reaches 2.43, in good agreement with the theoretical
prediction (see also \cite{Chow Bell}), and exceeds the classical
bound by up to 22 standard deviations when averaged over 10$^{6}$
sequences. 

\begin{figure}[h]
\includegraphics[width=8cm]{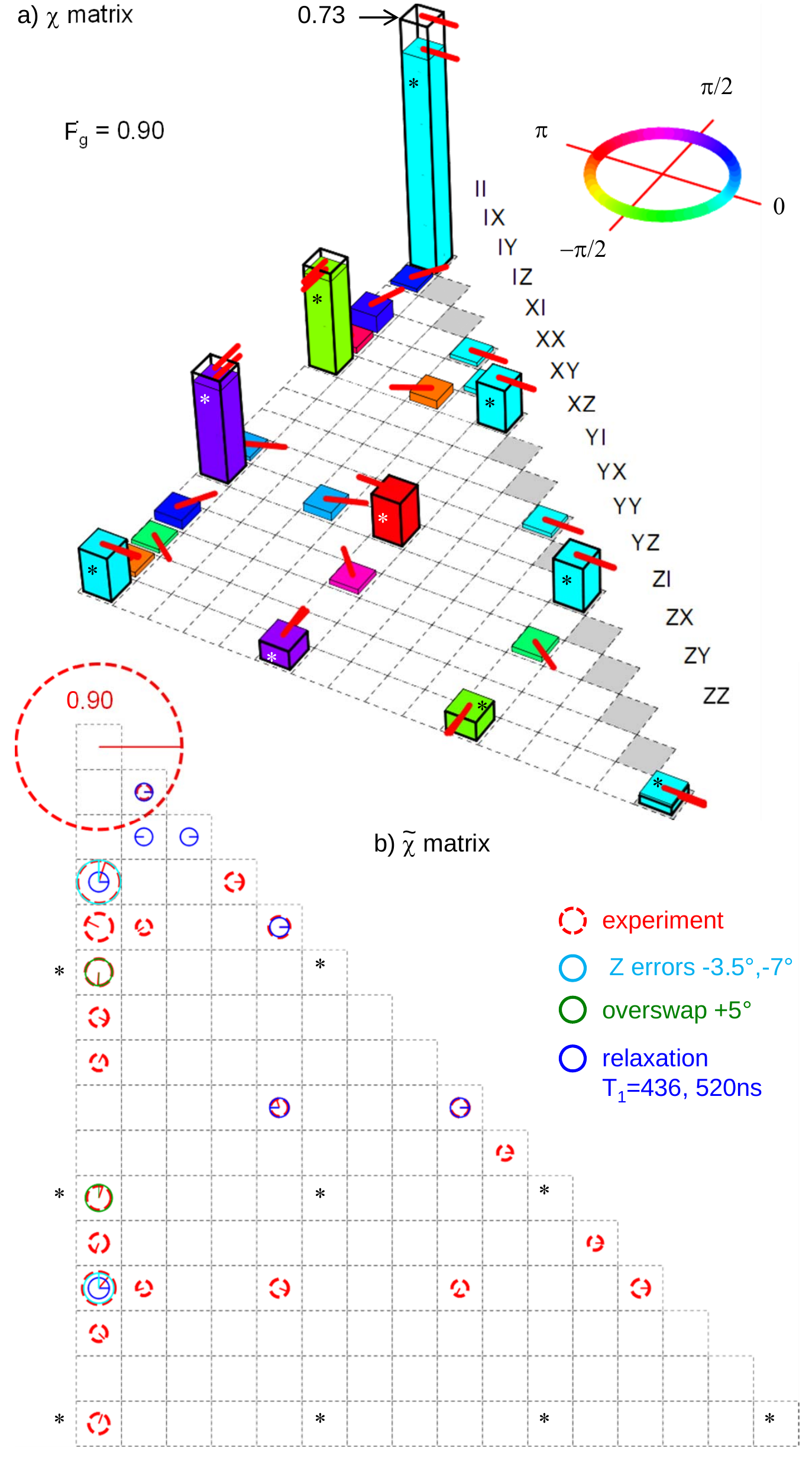}

\caption{Map of the implemented $\sqrt{iSWAP}$ gate yielding a fidelity of
90\%. (a) Superposition of the ideal (empty thick bars) and experimental
(color filled bars) lower part of the Hermitian matrix $\chi$ (elements
below 1\% not shown). Each complex matrix element is represented by
a bar with height proportional to its modulus and a red phase pointer
at the top of the bar (as well as a filling color for experiment)
giving its argument (top left inset). Expected peaks are marked by
a star. (b) Lower part of the $\widetilde{\chi}$ error matrix (red
circles - see text), with the same convention as in Fig.\ref{fig2},
but with circles magnified for readability (a one-cell diameter represents
a 8 \% modulus). Main visible contributions (continuous circles) are
explained in text.}

\label{fig4}%
\end{figure}

In a last experiment, we characterize the imperfections of our $\sqrt{iSWAP}$
gate by quantum process tomography \cite{Nielsen Chuang}. We build
a completely positive map $\rho_{\mathrm{out}}=\mathcal{E}(\rho_{\mathrm{in}})=\sum_{\mathrm{m,n}}\chi_{\mathrm{mn}}P_{\mathrm{m}}^{'}\rho_{\mathrm{in}}P_{\mathrm{n}}^{'\dagger}$
characterized by a $16\times16$ matrix $\chi$ expressed here in
the modified Pauli operator basis $\left\{ P_{\mathrm{k}}^{'}\right\} =\{I,X,Y^{'}=iY,Z\}^{\otimes2}$,
for which all matrices are real. For that purpose, we apply the gate
(\textcolor{black}{using pulse sequences similar to that of \ref{fig1}c,
with $\Delta t=31\,\mathrm{ns}$ and $\Theta_{\mathrm{j}}^{-1}$rotations})
to the sixteen input states $\{\left|0\right\rangle ,\left|1\right\rangle ,\left|0\right\rangle +\left|1\right\rangle ,\left|0\right\rangle +i\left|1\right\rangle \}^{\otimes2}$
and characterize both the input and output states by quantum state
tomography. By operating as described previously, we would obtain
apparent input and output density matrices including errors made in
the state tomography itself, which we don't want to include in the
gate map. Instead, we fit the 16 experimental input Pauli sets by
a model including amplitude and phase errors for the $X$ and $Y$
preparation and tomographic pulses (see S4), in order to determine
which operator set $\left\{ P_{\mathrm{k}}^{\mathrm{e}}\right\} $
is actually measured. The input and output matrices $\rho_{\mathrm{in,out}}$
corrected from the tomographic errors only are calculated by inverting
the linear relation $\left\{ \left\langle P_{\mathrm{k}}^{\mathrm{e}}\right\rangle =Tr\left(\rho P_{\mathrm{k}}^{\mathrm{e}}\right)\right\} $
and by applying it to the experimental Pauli sets. We then calculate
from the $\left\{ \rho_{\mathrm{in,out}}\right\} $ set an Hermitian
$\chi$ matrix that is not necessarily physical due to statistical
errors, and which we render physical by taking the nearest Hermitian
positive matrix. This final $\chi$ matrix is shown in Fig. \ref{fig4}
and compared to the ideal one, $\chi_{\mathrm{id}}$, which yields
a gate fidelity $F_{\mathrm{g}}=Tr\left(\chi.\chi_{\mathrm{id}}\right)=0.9$
\cite{Fidelities}. To better understand the imperfections, we also
show the map $\widetilde{\chi}$ of the actual process preceded by
the inverse ideal process \cite{Korotkov}. The first diagonal element
of $\widetilde{\chi}$ is equal to $F$ by construction. Then, main
visible errors arise from unitary operations and reduce fidelity by
1-2\% (a fit yields a too long coupling time inducing a 95\textdegree{}
swap instead of 90\textdegree{} and $\Theta_{1,2}$ rotations too
small by $3.5\lyxmathsym{\textdegree}$ and $7\lyxmathsym{\textdegree}$
respectively). On the other hand the known relaxation and dephasing
times reduces fidelity by 8\% but is barely visible in $\widetilde{\chi}$
due to a spread over many matrix elements with modulus of the order
of or below the $1-2\%$ noise level. 

As a conclusion, we have demonstrated a high fidelity $\sqrt{iSWAP}$
gate in a two josephson qubit circuit with individual non-destructive
single-shot readouts, observed a violation of the CHSH-Bell inequality,
and followed the register's dynamics by tomography. Although quantum
coherence and readout fidelity are still limited in this circuit,
they are sufficient to test in the near future simple quantum algorithms
and get their result in a single run, which would demonstrate the
concept of quantum speed-up.

We gratefully acknowledge discussions with J. Martinis and his coworkers,
with M. Devoret, D. DiVicenzo, A. Korotkov, P. Milman, and within
the Quantronics group, technical support from P. Orfila, P. Senat,
and J.C. Tack, as well as financial support from the European research
contracts MIDAS and SOLID, from ANR Masquelspec and C'Nano and from
the German Ministry of Education and Research.

\section*{Supplementary material }

S1. Sample preparation\\

The sample is fabricated on a silicon chip oxidized over 50 nm. A
150 nm thick niobium layer is first deposited by magnetron sputtering
and then dry-etched in a $SF_{6}$ plasma to pattern the readout resonators,
the current lines for frequency tuning, and their ports. Finally,
the transmon qubit, the coupling capacitance and the Josephson junctions
of the resonators are fabricated by double-angle evaporation of aluminum
through a shadow mask patterned by e-beam lithography. The first layer
of aluminum is oxidized in a $Ar-O_{2}$ mixture to form the oxide
barrier of the junctions. The chip is glued with wax on a printed
circuit board (PCB) and wire bonded to it. The PCB is then screwed
in a copper box anchored to the cold plate of a dilution refrigerator.\\

S2. Sample parameters\\

The sample is first characterized by spectroscopy (see Fig.$\,$1.b
of main text). The incident power used is high enough to observe the
resonator frequency $\nu_{\mathrm{R}}$, the qubit line $\nu_{01}$,
and the two-photon transition at frequency $\nu_{02}/2$ between the
ground and second excited states of each transmon (data not shown).
A fit of the transmon model to the data yields the sample parameters
$E_{\mathrm{J}}^{\mathrm{I}}/h=36.2\,\mathrm{GHz}$, $E_{\mathrm{C}}^{\mathrm{I}}/h=0.98\,\mathrm{GHz}$,
$d_{I}=0.2$, $E_{\mathrm{J}}^{\mathrm{II}}/h=43.1\,\mathrm{GHz}$,
$E_{\mathrm{C}}^{\mathrm{II}}/h=0.87\,\mathrm{GHz}$, $d_{\mathrm{II}}=0.35$,
$\nu_{\mathrm{R}}^{\mathrm{I}}=6.84\,\mathrm{GHz}$, and $\nu_{\mathrm{R}}^{\mathrm{II}}=6.70\,\mathrm{GHz}$.
The qubit-readout anticrossing at $\nu=\nu_{\mathrm{R}}$ yields the
qubit-readout couplings $g_{0}^{\mathrm{I}}\simeq g_{0}^{\mathrm{II}}\simeq50\,\mathrm{MHz}$.
Independent measurements of the resonator dynamics (data not shown)
yield quality factors $Q_{\mathrm{I}}=Q_{\mathrm{II}}=730$ and Kerr
non linearities \cite{JBA Mallet,FlorianKerr} $K_{\mathrm{I}}/\nu_{\mathrm{R}}^{\mathrm{I}}\simeq K_{\mathrm{II}}/\mathrm{\nu}_{\mathrm{R}}^{II}\simeq-2.3\pm0.5\times10^{-5}$.\\

S3. Experimental setup
\begin{itemize}
\item Qubit microwave pulses: The qubit drive pulses are generated by two
phase-locked microwave generators whose continuous wave outputs are
fed to a pair of I/Q-mixers. The two IF inputs of each of these mixers
are provided by a 4-Channel$1\,\mathrm{GS/s}$ arbitrary waveform
generator (AWG Tektronix AWG5014). Single-sideband mixing in the frequency
range of 50-300 MHz is used to generate multi-tone drive pulses and
to obtain a high ON/OFF ratio ($>\,50\,\mathrm{dB}$) of the signal
at the output of the mixers. Phase and amplitude errors of the mixers
are corrected by measuring the signals at the output and applying
sideband and carrier frequency dependent corrections in amplitude
and offset to the IF input channels. 
\item Flux Pulses: The flux control pulses are generated by a second AWG
and sent to the chip through a transmission line, equipped with 40
dB of attenuation distributed over different temperature stages and
a pair of 1 GHz absorptive low-pass filters at $4\,\mathrm{K}$. The
input signal of each flux line is fed back to room temperature through
an identical transmission line and measured to compensate the non-ideal
frequency response of the line.
\item Readout Pulses: The pulses for the Josephson bifurcation amplifier
(JBA) readouts are generated by mixing the continuous signals of a
pair of microwave generators with IF pulses provided by a $1\,\mathrm{GS/s}$
arbitrary function generator. Each readout pulse consists of a measurement
part with a rise time of $30\,\mathrm{ns}$ and a hold time of 100
ns, followed by a $2\,\mu s$ long latching part at 90 \% of the pulse
height. 
\item Drive and Measurement Lines: The drive and readout microwave signals
of each qubit are combined and sent to the sample through a pair of
transmission lines that are attenuated by 70 dB over different temperature
stages and filtered at $4\mathrm{\, K}$ and $300\,\mathrm{mK}$.
A microwave circulator at $20\,\mathrm{mK}$ separates the input signals
going to the chip from the reflected signals coming from the chip.
The latter are amplified by $36\,\mathrm{dB}$ at $4\,\mathrm{K}$
by two cryogenic HEMT amplifiers (CIT Cryo 1) with noise temperature
$5\,\mathrm{K}$. The reflected readout pulses get further amplified
at room temperature and demodulated with the continuous signals of
the readout microwave sources. The IQ quadratures of the demodulated
signals are sampled at $1\,\mathrm{GS/s}$ by a 4-channel Data Acquisition
system (Acqiris DC282). 
\end{itemize}
S4. Readout characterization\\

\begin{figure}[tbph]
\includegraphics[width=12cm]{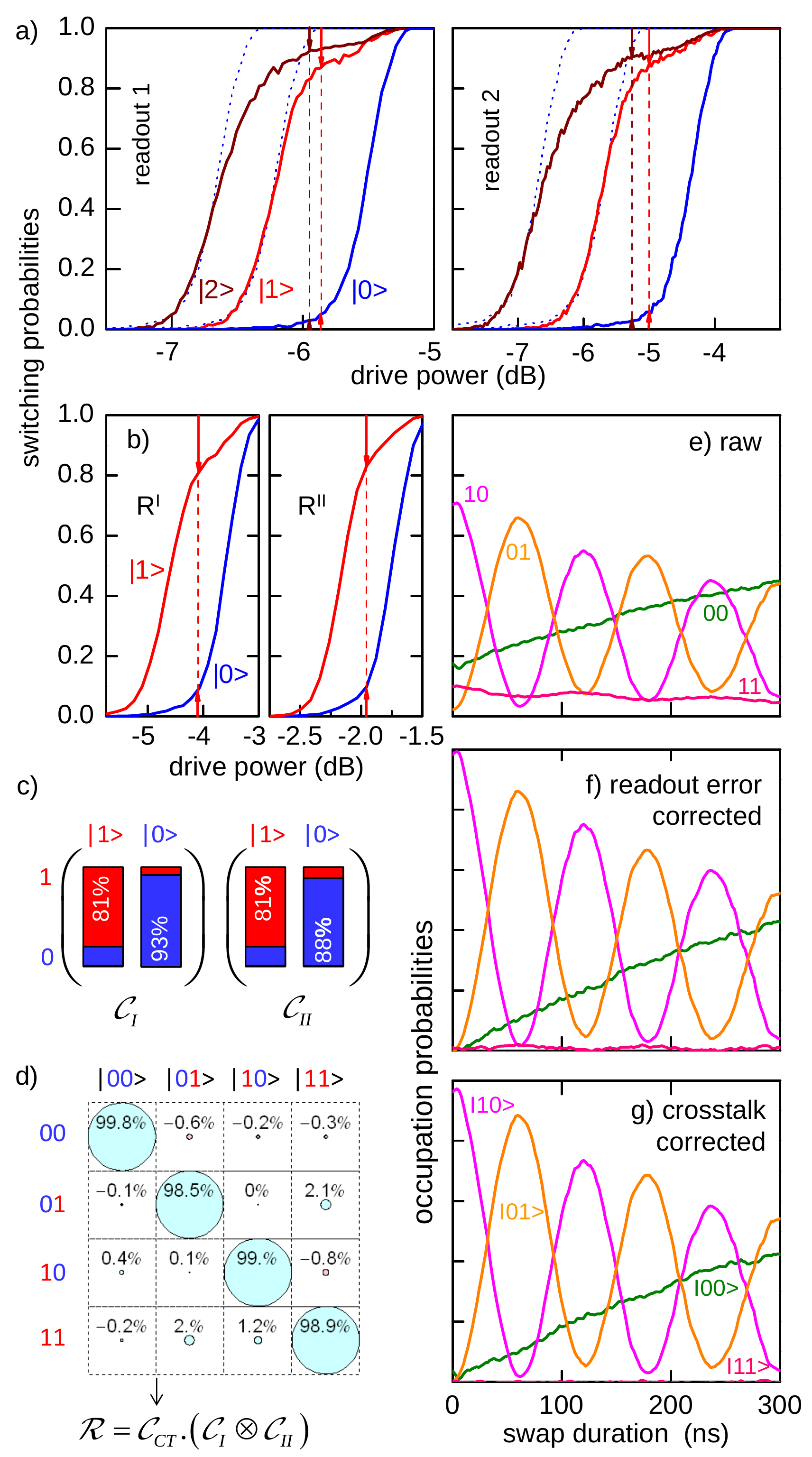}

\caption{Readout imperfections and their correction. (a) Switching probabilities
of the readouts as a function of their driving power, with the qubit
prepared in state$\left|0\right\rangle $ (blue), $\left|1\right\rangle $
( red), or $\left|2\right\rangle $ (brown), at the optimal readout
points. The arrows and dashed segments indicate the readout errors
and contrast, at the power where the later is maximum. (b) Same as
(a) but at readout points $R^{\mathrm{I,II}}$ used in this work.
(c-d) Single readout matrices $\mathcal{C}_{\mathrm{I,II}}$ and pure
readout crosstalk matrix $\mathcal{C}_{\mathrm{CT}}$ characterizing
the simultaneous readout of the two qubits. (e-g) bare readout outcomes
$uv$, outcomes corrected from the independent readout errors only,
and$\left|uv\right\rangle $ population calculated with the full correction
including crosstalk for the swapping experiment of Fig.\ref{fig2}.}

\label{fig S4.1}%
\end{figure}

Errors in our readout scheme are discussed in detail in {[}13{]} for
a single qubit. First, incorrect mapping $\left|0\right\rangle \rightarrow1$
or$\left|1\right\rangle \rightarrow0$ of the projected state of the
qubit to the dynamical state of the resonator can occur, due to the
stochastic nature of the switching between the two dynamical states.
As shown in Fig.\ref{fig S4.1}, the probability $p$ to obtain the
outcome 1 varies continuously from 0 to 1 over a certain range of
drive power $P_{\mathrm{d}}$ applied to the readout. When the shift
in power between the two $p_{\left|0\right\rangle ,\left|1\right\rangle }(P_{\mathrm{d}})$
curves is not much larger than this range, the two curves overlap
and errors are significant even at the optimal drive power where the
difference in $p$ is maximum. Second, even in the case of non overlapping
$p_{\left|0\right\rangle ,\left|1\right\rangle }(P_{\mathrm{d}})$
curves, the qubit initially projected in state$\left|1\right\rangle $
can relax down to $\left|0\right\rangle $ before the end of the measurement,
yielding an outcome 0 instead of 1. The probability of these two types
of errors vary in opposite directions as a function of the frequency
detuning $\Delta=\nu_{\mathrm{R}}-\nu>0$ between the resonator and
the qubit, so that a compromise has to be found for $\Delta$. Besides,
the contrast $c=Max\left(p_{\left|1\right\rangle }-p_{\left|0\right\rangle }\right)$
can be increased {[}12{]} by shelving state $\left|1\right\rangle $
into state $\left|2\right\rangle $ with a microwave $\pi$ pulse
at frequency $\nu_{12}$ just before the readout resonator pulse.
The smallest errors $e_{0}^{\mathrm{I,II}}$ and $e_{1}^{\mathrm{I,II}}$
when reading $\left|0\right\rangle $ and $\left|1\right\rangle $
are found for $\Delta_{\mathrm{I}}=440\,\mathrm{MHz}$ and $\Delta_{\mathrm{II}}=575\,\mathrm{MHz}$
and are shown by arrows in the top panels of Fig.\ref{fig S4.1}:
$e_{0}^{\mathrm{I}}=5\%$ and $e_{1}^{\mathrm{I}}=13\%$ (contrast
$c_{\mathrm{I}}=1-e_{0}^{\mathrm{I}}-e_{1}^{\mathrm{I}}=82\%$), and
$e_{0}^{\mathrm{II}}=5.5\%$ and $e_{1}^{\mathrm{II}}=12\%$ ($c_{\mathrm{II}}=82\%$).
When using the $\left|1\right\rangle \rightarrow\left|2\right\rangle $
shelving before readout, $e_{0}^{\mathrm{I}}=2.5\%$ and $e_{2}^{\mathrm{I}}=9.5\%$
(contrast $c_{\mathrm{I}}==1-e_{0}^{\mathrm{I}}-e_{2}^{\mathrm{I}}=88\%$),
and $e_{0}^{\mathrm{II}}=3\%$ and $e_{2}^{\mathrm{II}}=8\%$ ($c_{\mathrm{II}}=89\%$).
These best results are very close to those obtained in {[}12{]}, but
are unfortunately not relevant to this work.

Indeed, when the two qubits are measured simultaneously, one has also
to take into account a possible readout crosstalk, i.e. an influence
of the projected state of each qubit on the outcome of the readout
of the other qubit. We do observe such an effect and have to minimize
it by increasing $\Delta_{\mathrm{I,II}}$ up to $\sim1\,\mathrm{GHz}$
with respect to previous optimal values and by not using the shelving
technique. An immediate consequence shown in Fig.\ref{fig S4.1}(b)
is a reduction of the $c_{\mathrm{I,II}}$ contrasts. The errors when
reading $\left|0\right\rangle $ and $\left|1\right\rangle $ are
now $e_{0}^{\mathrm{I}}=19\,\%$ and $e_{1}^{\mathrm{I}}=7\,\%$ (contrast
$c_{\mathrm{I}}=74\%$) and $e_{0}^{\mathrm{II}}=19\,\%$ and $e_{1}^{\mathrm{II}}=12\,\%$
(contrast $c_{\mathrm{II}}=69\%$). Then to characterize the errors
due to crosstalk, we measure the $4\times4$ readout matrix $\mathcal{R}$
linking the probabilities $p_{\mathrm{uv}}$ of the four possible
$uv$ outcomes to the population of the four $\left|uv\right\rangle $
states. As shown in Fig.\ref{fig S4.1}(c-d), we then rewrite $\mathcal{R}=\mathcal{C}_{\mathrm{CT}}.\left(\mathcal{C}_{\mathrm{I}}\otimes\mathcal{C}_{\mathrm{II}}\right)$
as the product of a $4\times4$ pure crosstalk matrix $\mathcal{C}_{\mathrm{CT}}$
with the tensorial product of the two $2\times2$ single qubit readout
matrices \[
\mathcal{C}_{I,II}=\left(\begin{array}{cc}
1-e_{0}^{\mathrm{I,II}} & e_{1}^{\mathrm{I,II}}\\
e_{0}^{\mathrm{I,II}} & 1-e_{1}^{\mathrm{I,II}}\end{array}\right).\]

We also illustrate on the figure the impact of the readout errors
on our swapping experiment by comparing the bare readout outcomes
$uv$, the outcomes corrected from the independent readout errors
only, and the$\left|uv\right\rangle $ population calculated with
the full correction including crosstalk.

We now explain briefly the cause of the readout crosstalk in our processor.
Unlike what was observed for other qubit readout schemes using switching
detectors {[}5{]}, the crosstalk we observe is not directly due to
an electromagnetic perturbation induced by the switching of one detector
that would help or prevent the switching of the other one. Indeed,
when both qubits frequencies $\nu_{\mathrm{I,II}}$ are moved far
below $\nu_{\mathrm{R}}^{\mathrm{I,II}}$, the readout crosstalk disappears:
the switching of a detector has no measurable effect on the switching
of the other one. The crosstalk is actually due to the rather strong
ac-Stark shift $\sim2\left(n_{\mathrm{H}}-n_{\mathrm{L}}\right)g_{0}^{2}/(R-\nu_{\mathrm{R}})\sim500\,\mathrm{MHz}$
of the qubit frequency when a readout resonator switches from its
low to high amplitude dynamical state with $n_{\mathrm{L}}\sim10$
and $n_{\mathrm{H}}\sim10^{2}$ photons, respectively. The small residual
effective coupling between the qubits at readout can then slightly
shift the frequency of the other resonatator, yielding a change of
its switching probability by a few percent. Note that coupling the
two qubits by a resonator rather than by a fixed capacitor would solve
this problem.\\

S5. Removing errors on tomographic pulses before calculating the gate
process map\\

\begin{figure}[tbph]
\includegraphics[width=17cm]{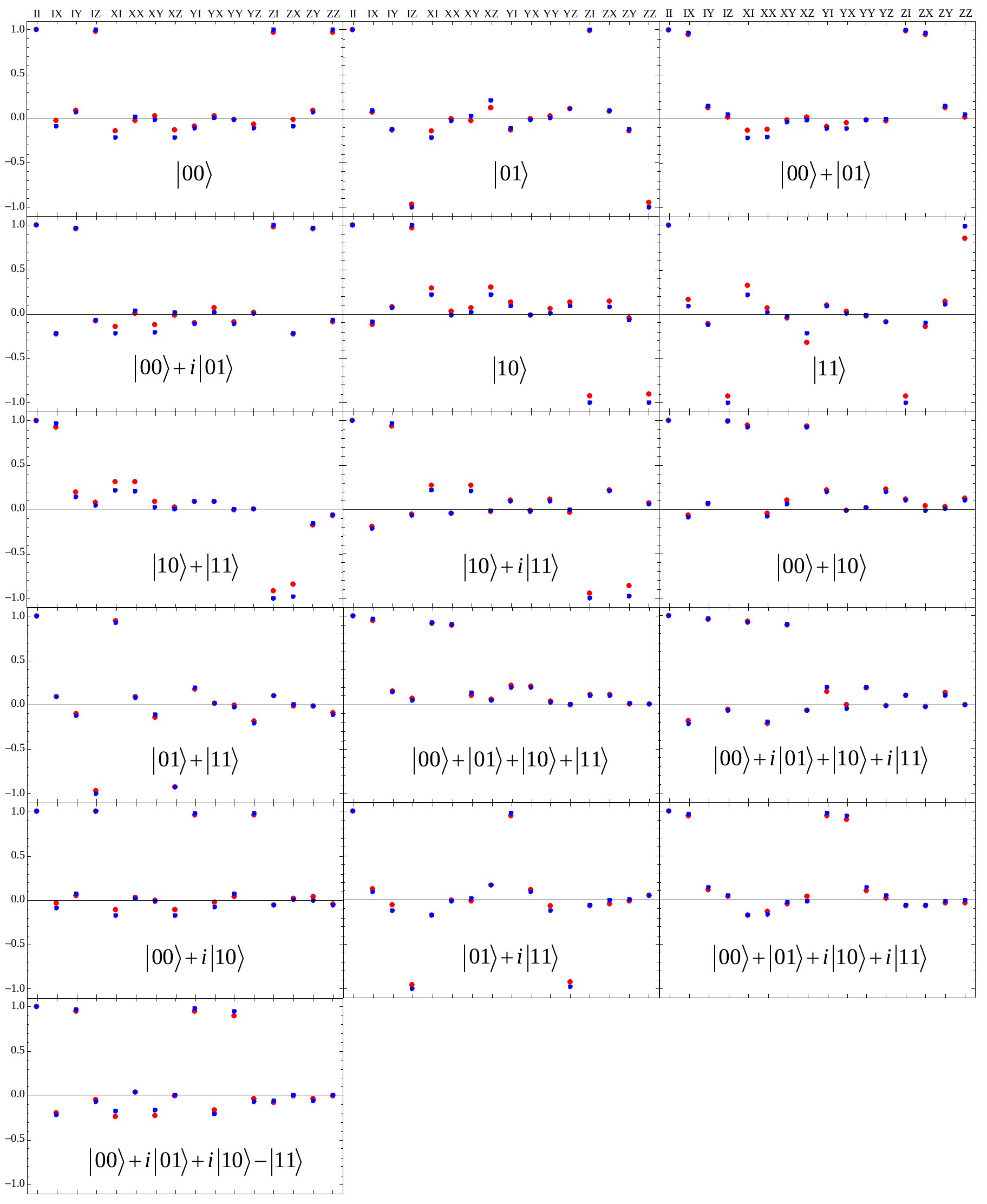}

\caption{Fitting of the pulse errors at state preparation and tomography. Measured
(red) and fitted (blue - see text) Pauli sets $\left\langle P_{\mathrm{k}}^{\mathrm{e}}\right\rangle $
for the sixteen targeted input states $\{\left|0\right\rangle ,\left|1\right\rangle ,\left|0\right\rangle +\left|1\right\rangle ,\left|0\right\rangle +i\left|1\right\rangle \}^{\otimes2}$.
The $\{II,IX,IY,IZ,XI,...\}$ operators indicated in abscisse are
the targeted operators and not those actually measured (due to tomographic
errors).}

\label{fig S5.1}%
\end{figure}
Tomographic errors are removed from the process map of our $\sqrt{iSWAP}$
gate using the following method. The measured Pauli sets corresponding
to the sixteen input states are first fitted by a model including
errors both in the preparation of the state (index $prep$) and in
the tomographic pulses (index $tomo$). The errors included are angular
errors $\varepsilon_{\mathrm{I,II}}^{\mathrm{prep}}$ on the nominal
$\pi$ rotations around $X_{\mathrm{I,II}}$, $\eta_{\mathrm{I,II}}^{\mathrm{prep,tomo}}$and
$\delta_{\mathrm{I,II}}^{\mathrm{prep,tomo}}$ on the nominal $\pi/2$
rotations around $X_{\mathrm{I,II}}$ and $Y_{\mathrm{I,II}}$, a
possible departure $\xi_{\mathrm{I,II}}$ from orthogonality of $\left(\overrightarrow{X_{\mathrm{I}}},\overrightarrow{Y_{\mathrm{I}}}\right)$
and $\left(\overrightarrow{X_{\mathrm{II}}},\overrightarrow{Y_{\mathrm{II}}}\right)$,
and a possible rotation $\mu_{\mathrm{I,II}}$ of the tomographic
$XY$ frame with respect to the preparation one. The rotation operators
used for preparing the states and doing their tomography are thus
given by

\[
\begin{array}{c}
X_{\mathrm{I,II}}^{\mathrm{prep}}(\pi)=e^{-\mathrm{i}\left(\pi+\varepsilon_{\mathrm{I,II}}^{\mathrm{prep}}\right)\sigma_{\mathrm{x}}^{\mathrm{I,II}}/2},\\
X_{\mathrm{I,II}}^{\mathrm{prep}}(-\pi/2)=e^{+\mathrm{i}\left(\pi/2+\eta_{\mathrm{I,II}}^{\mathrm{prep}}\right)\sigma_{\mathrm{x}}^{\mathrm{I,II}}/2},\\
Y_{\mathrm{I,II}}^{\mathrm{prep}}(\pi/2)=e^{-\mathrm{i}\left(\pi/2+\delta_{\mathrm{I,II}}^{\mathrm{prep}}\right)\left[\mathrm{cos}\left(\xi_{\mathrm{I,II}}\right)\sigma_{\mathrm{y}}^{\mathrm{I,II}}\mathrm{-sin}\left(\xi_{\mathrm{I,II}}\right)\sigma_{\mathrm{x}}^{\mathrm{I,II}}\right]/2},\\
X_{\mathrm{I,II}}^{\mathrm{tomo}}(\pi/2)=e^{-\mathrm{i}\left(\pi/2+\eta_{\mathrm{I,II}}^{\mathrm{tomo}}\right)\left[\mathrm{\mathrm{sin}\left(\mu_{I,II}\right)\sigma_{x}^{I,II}+cos}\left(\mu_{\mathrm{I,II}}\right)\sigma_{\mathrm{y}}^{\mathrm{I,II}}\right]/2},\\
Y_{\mathrm{I,II}}^{\mathrm{tomo}}(-\pi/2)=e^{+\mathrm{i}\left(\pi/2+\delta_{\mathrm{I,II}}^{\mathrm{tomo}}\right)\left[\mathrm{cos}\left(\mu_{\mathrm{I,II}}+\xi_{\mathrm{I,II}}\right)\sigma_{\mathrm{y}}^{\mathrm{I,II}}\mathrm{-sin}\left(\mu_{\mathrm{I,II}}+\xi_{\mathrm{I,II}}\right)\sigma_{x}^{\mathrm{I,II}}\right]/2}.\end{array}\]
The sixteen input states are then $\left\{ \rho_{\mathrm{in}}^{\mathrm{e}}=U\left|0\right\rangle \left\langle 0\right|U^{\dagger}\right\} $
with $\left\{ U\right\} =\{I_{\mathrm{I}},X_{\mathrm{I}}^{\mathrm{prep}}(\pi),Y_{\mathrm{I}}^{\mathrm{prep}}(\pi/2),X_{\mathrm{I}}^{\mathrm{prep}}(-\pi/2)\}\otimes\{I_{\mathrm{II}},X_{\mathrm{II}}^{\mathrm{prep}}(\pi),Y_{\mathrm{II}}^{\mathrm{prep}}(\pi/2),X_{\mathrm{II}}^{\mathrm{prep}}(-\pi/2)\}$,
and each input state yields a Pauli set $\left\{ \left\langle P_{\mathrm{k}}^{\mathrm{e}}\right\rangle =Tr\left(\rho_{\mathrm{in}}^{\mathrm{e}}P_{\mathrm{k}}^{\mathrm{e}}\right)\right\} $
with $\left\{ P_{\mathrm{k}}^{\mathrm{e}}\right\} =\{I_{\mathrm{I}},X_{\mathrm{I}}^{\mathrm{e}},Y_{\mathrm{I}}^{\mathrm{e}},Z_{\mathrm{I}}\}\otimes\{I_{\mathrm{II}},X_{\mathrm{II}}^{\mathrm{e}},Y_{\mathrm{II}}^{\mathrm{e}},Z_{\mathrm{II}}\}$,
$X^{\mathrm{e}}=Y^{\mathrm{tomo}}(-\pi/2)^{\dagger}\sigma_{z}Y^{\mathrm{tomo}}(-\pi/2)$,
and $Y^{\mathrm{e}}=X^{\mathrm{tomo}}(\pi/2)^{\dagger}\sigma_{\mathrm{z}}X^{\mathrm{tomo}}(\pi/2)$.
Figure \ref{fig S5.1} shows the best fit of the modelled $\left\{ \left\langle P_{k}^{e}\right\rangle \right\} $
set to the measured input Pauli sets, yielding $\varepsilon_{\mathrm{I}}^{\mathrm{prep}}=-1\text{\textdegree}$,
$\varepsilon_{\mathrm{II}}^{\mathrm{prep}}=-3\text{\textdegree}$,
$\eta_{\mathrm{I}}^{\mathrm{prep}}=3\text{\textdegree}$, $\mathrm{\eta}_{\mathrm{II}}^{\mathrm{prep}}=4\text{\textdegree}$,
$\delta_{\mathrm{I}}^{\mathrm{prep}}=-6\text{\textdegree}$, $\delta_{\mathrm{II}}^{\mathrm{prep}}=-3\text{\textdegree}$,
$\eta_{\mathrm{I}}^{\mathrm{tomo}}=-6\text{\textdegree}$, $\eta_{\mathrm{II}}^{\mathrm{tomo}}=-4\text{\textdegree}$,
$\lambda_{\mathrm{I}}^{t\mathrm{omo}}=12\text{\textdegree}$, $\lambda_{\mathrm{II}}^{\mathrm{tomo}}=5\text{\textdegree}$,
$\xi_{\mathrm{I}}=1\text{\textdegree}$, $\xi_{\mathrm{II}}=-2\text{\textdegree}$,
and $\mu_{\mathrm{I}}=\mu_{\mathrm{II}}=-11\text{\textdegree}$.

Knowing the tomographic errors and thus $\left\{ \left\langle P_{\mathrm{k}}^{\mathrm{e}}\right\rangle \right\} $,
we then invert the linear relation $\left\{ \left\langle P_{\mathrm{k}}^{\mathrm{e}}\right\rangle =Tr\left(\rho P_{\mathrm{k}}^{\mathrm{e}}\right)\right\} $
to find the $16\times16$ matrix $B$ that links the vector $\overrightarrow{\left\langle P_{\mathrm{k}}^{\mathrm{e}}\right\rangle }$
to the columnized density matrix $\overrightarrow{\rho}$, i.e. $\overrightarrow{\rho}=B.\overrightarrow{\left\langle P_{\mathrm{k}}^{\mathrm{e}}\right\rangle }$.
The matrix $B$ is finally applied to the measured sixteen input and
sixteen output Pauli sets to find the sixteen $(\rho_{\mathrm{in},},\rho_{\mathrm{out}})_{\mathrm{k}}$
couples to be used for calculating the gate map.

\begin{thebibliography}{17}
\bibitem{Fidelities} Note that $F_{g}$ is also equal to Shumacher's
fidelity $Tr\left[S_{id}^{\dagger}S\right]/Tr\left[S_{id}^{\dagger}S_{id}\right]$
with $S$ (resp. $S_{id}$) the super operator of the actual (resp.
ideal unitary) process, and that fidelities $F$ for the 16 outputs
states range between 80\% and 99.5\%
\end{thebibliography}
\end{document}